\documentclass[aps,pra,twocolumn,showpacs,amsmath,amssymb]{revtex4}
\usepackage{graphicx}
\usepackage{bm}

\begin{document}
\title{Transistor-Like Behavior of a Bose-Einstein Condensate in a Triple Well Potential.}
\author{James A. Stickney}
\affiliation{Department of Physics, Worcester Polytechnic Institute,
100 Institute Road, Worcester, Massachusetts 01609}
\author{Dana Z. Anderson}
\affiliation{Department of Physics and JILA, University of Colorado
and National Institute of Standards and Technology, Boulder,
Colorado 80309-0440}
\author{Alex A. Zozulya}
\affiliation{Department of Physics, Worcester Polytechnic Institute,
100 Institute Road, Worcester, Massachusetts 01609} \email[]{
zozulya@wpi.edu}

\begin{abstract}
In the last several years considerable efforts have been devoted to
developing Bose-Einstein Condensate (BEC)-based devices for
applications such as fundamental research, precision measurements
and integrated atom optics. Such devices capable of complex
functionality can be designed from simpler building blocks as is
done in microelectronics. One of the most important components of
microelectronics is a transistor. We demonstrate that Bose-Einstein
condensate in a three well potential structure where the tunneling
of atoms between two wells is controlled by the population in the
third, shows behavior similar to that of an electronic field effect
transistor. Namely, it exhibits switching and both absolute and
differential gain. The role of quantum fluctuations is analyzed,
estimates of switching time and parameters for the potential are
presented.
\end{abstract}



\maketitle

\section{Introduction}\label{sec:intro}

Recent experimental realizations of atom optical devices such as
atomic waveguides, beamsplitters
\cite{muller99,dekker00,cassetarri00,leanhardt02}, on-chip BEC
sources and conveyor belts \cite{hanselnature01,hommelhoff05}
opens a way for development of more complex devices such as, e.g.,
BEC-based interferometers \cite{wang05,shin04}. On-chip integrated
cold atom circuits capable of complex functionality can be
constructed from simpler building blocks as it is done in
microelectronics to find applications in fundamental physics,
precision measurements and quantum information technology.

One of the most important components of a microelectronic circuit
is a transistor. In this paper we present a BEC-based device which
will be subsequently called a BEC transistor or an atom
transistor. It enables one to control a large number of atoms with
a smaller number of atoms and demonstrates switching and both
differential and absolute gain thus showing behavior similar to
that of an electronic transistor. The device is not optimized for
performance but is arguably the simplest possible geometry showing
behavior reminiscent of a transistor. This makes its experimental
realization relatively easy with existing atom chip techniques.

\begin{figure}
\includegraphics[width=8.6cm]{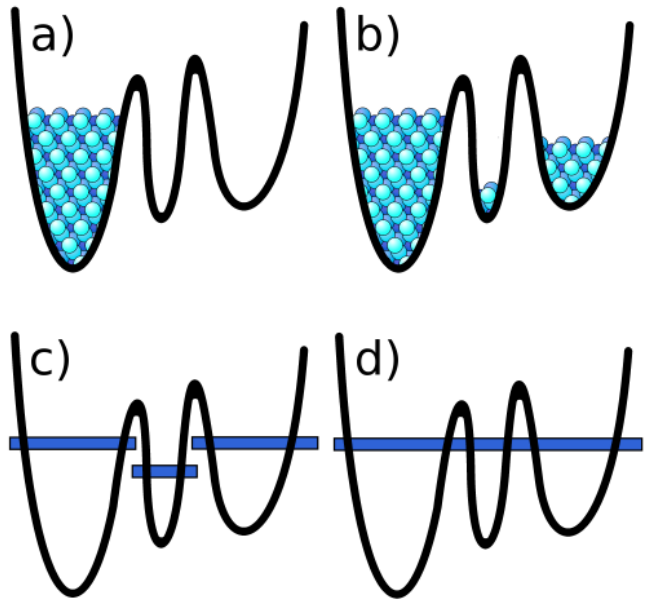}%
\caption{\label{fig:toon} The geometry of a BEC transistor. When the
number of atoms in the middle well is small, tunneling from the left
into the right well is negligible (a). This is due to the fact that
the chemical potential of the middle well does not match that of the
two other wells (c).  Placing atoms in the middle well increases the
chemical potential due to interatomic interactions (d) and enables
tunneling then atoms tunnel from the left into the right well. This
happpens because atom-atom interactions increase the energy of the
middle guide (b). }
\end{figure}

The BEC transistor uses a Bose Einstein condensate in a triple well
potential, as shown schematically in Fig \ref{fig:toon}. In fact,
Fig.~\ref{fig:toon} refers to two subtly different possible
experimental realization of the device. In the trapped
configuration, the BEC is confined in all three dimensions in the
potential wells. The wells are allowed to interact for time interval
$T$. This is done either spatially bringing them together and
separating apart after time $T$ of changing the shapes of the
potential wells so that the interaction is suppressed after time
$T$. In the waveguide configuration, the potential wells of
Fig.~\ref{fig:toon} represent three guides that converge, run
parallel to each other for distance $L$ and then diverge. The
interaction time $T = L/v$ in this geometry is determined by the
speed of flow $v$ of the BEC in the guides. In the following for
definiteness we will use the terminology appropriate for the trapped
configuration.

The BEC transistor is similar to an electronic field effect
transistor. The left well behaves like the source, the middle as the
gate, and the right well is equivalent to the drain. If there are no
atoms in the middle well, practically no atoms tunnel from the left
into the right well, as shown in Fig.~\ref{fig:toon}a. A small
number of atoms placed into the middle well switches the device
resulting into the strong flux of atoms from the left well (the
source) through the middle and into the right well as shown in
Fig.~\ref{fig:toon}b. Increasing number of atoms in the middle well
increases the number of atoms that tunnel into the right well.
Parameters of the triple well structure are chosen so that the
number of atoms having tunneled into the right well at the end of
the interaction period is much larger that the number of atoms in
the middle well. In the subsequent sections we will show that the
BEC transistor exhibits both absolute and differential gain.

The physics of operation of the BEC transistor is based on atom-atom
interactions and appropriate design of the potentials. The chemical
potential of the left well is chosen to be nearly equal to the
ground state energy level of the empty right well (in
Fig.~\ref{fig:toon} we make them equal). The ground state energy of
the empty middle well is chosen to be considerably lower than that
in both the left and the right wells. Placing atoms in a well raises
the value of chemical potential due to atom-atom interactions.
Parameters of the potential wells are chosen so that the chemical
potential in the middle well is considerably more sensitive to the
change in the number of atoms in the well than is the case for the
left and right wells.

When the middle and right wells are initially unpopulated, tunneling
of atoms from the left to the middle well is blocked because of the
energy mismatch as shown in Fig.~\ref{fig:toon}c. If some amount of
atoms is placed into the middle well, the atom-atom interactions
will increase the energy of the atoms in the middle well.  When the
chemical potential in the middle well becomes nearly equal to that
in the left and right wells, the device switches and atoms become
able to tunnel from the left through the middle into the right well
as shown in Fig. \ref{fig:toon}d.

Using atom-atom interactions to block tunneling in a double-well
structure is often referred to as self trapping. This effect was
first described in Ref.~\cite{milburn97}. If a condensate is placed
in one of the two weakly-coupled spatially separated potential wells
with matched energy levels, it can oscillate between the wells by
linear quantum tunneling. However, due to atom-atom interactions,
the tunneling is blocked when the number of atoms in the condensate
exceeds some critical value. This suppression is due to the fact
that interactions increase chemical potential of the atoms in the
occupied wells and introduce nonlinear energy mismatch. Self
trapping has been analyzed for a large number of systems including
asymmetric double well potentials \cite{smerzi97} and symmetric
three well systems \cite{franzosi01}.  It has also been observed
experimentally for atoms in a one dimensional optical lattice
\cite{anker05}.

The quantum state of two trapped Bose-Einstein condensates in a
double well potential has been analyzed in Ref. \cite{steel98}. It
has been shown that when the two wells are separated and the
interaction between the atoms is repulsive, the lowest energy
state is fragmented, which means that the coherence between the
atoms in each well is lost. The dependence of this fragmentation
on the splitting rate and physical parameters of the potential has
been analyzed in Refs.~\cite{spekkens99} and \cite{javanainen99}.
The visibility of interference fringes after splitting of a
condensate with both attractive and repulsive interactions was
analyzed in Ref.~\cite{jaaskelainen04}, who showed a decrease in
quantum noise in the case of attractive interactions.  The quantum
dynamics of atoms in a symmetric double well potential, where the
atoms are in an initially fragmented state was also analyzed in
\cite{jaaskelainen05}.

Bose Einstein condensates in triple well structures have been
analyzed and the stationary solutions in the mean field
approximation were found in Ref. \cite{nemoto00}.  Three well
systems show chaotic solutions \cite{franzosi03} and the dynamics of
atoms in a three well potential is sensitive to the initial
conditions of the system \cite{franzosi01}.  This means that one can
control the dynamics of the system not only by varying the physical
parameters of the potential, but also by changing the initial
conditions.

The authors of Ref.~\cite{micheli04} have recently proposed a single
atom transistor in a 1D optical lattice. A quantum interference
phenomenon is used to switch the flux of atoms in a lattice through
a cite containing a single impurity atom. Finally,
Ref.~\cite{seaman06} discusses ``atomtronic" diodes and transistors
which are direct analogs of their electronic counterparts.  These
devices use cold atoms in an optical lattice instead of electrons in
a crystal. With the addition of impurities both N type and P type
``semiconductors'' may be constructed.  A bipolar junction
transistor could be then built with a NPN or PNP sandwich.

The rest of the paper is organized as follows. Section
\ref{sec:eqns_of_motion} contains derivation of the general
equations of motion for a BEC in an n-well potential with arbitrary
shapes of the wells and discussion of the limits of validity of the
model. In Sec.~\ref{sec:3well_structure} we specialize our
discussion to the case of a three well structure.
Sec.~\ref{subsec:meanfild} is devoted to the analysis of the
equations of motion in the mean field limit and in Sec.
\ref{subsec:SQ} we will compare the results of the mean field to a
second quantization calculation. Section \ref{sec:discussion}
contains estimates of the physical parameters for the device and
discusses the possibility of its experimental realization.

\section{Equations of motion}\label{sec:eqns_of_motion}

The Hamiltonian for a system of interacting bosons in an external
potential $V(x)$ is of the form
\begin{equation}
    H = \int dx \hat \Psi^\dag \left[ -\frac{\hbar^2 }{ 2 m} \nabla^2
    + V(x) \right] \hat \Psi + \frac{U_0 }{ 2} \int d x \hat \Psi^\dag
    \hat \Psi^\dag \hat \Psi \hat \Psi. \label{eqn:SQ_H1}
\end{equation}
Here $\hat \Psi$ is the field operator and $U_0 = 4 \pi a_s
\hbar^{2}/ m$, where $m$ is the atomic mass and $a_s$ is the s-wave
scattering length. For notational simplicity we are considering
one-dimensional case. Extension to two or three dimensions is
straightforward.

In the standard basis of eigenfunctions $\psi_{i}$ of the linear
part of the Hamiltonian

\begin{equation}
    \left[ -\frac{\hbar^2 }{ 2 m} \nabla^2 + V(x) \right]\psi_{i} = \hbar
    \Omega_{i}\psi_{i},
    \label{eqn:SQ_H2}
\end{equation}
the field operator is represented as
\begin{equation}
    \hat \Psi = \sum_i \psi_i a_i, \label{eqn:decomp}
\end{equation}
where $a_{i}$ is the destruction operator for the mode $\psi_{i}$.
These operators satisfy the canonical commutation relations
\begin{eqnarray}
    &&[a_i, a_j^\dag] = \delta_{ij}, \nonumber \\
    &&[a_i, a_j] = 0. \label{eqn:commutator_a}
\end{eqnarray}
The potential $V(x)$ consists of $n$ weakly coupled potential
wells. The eigenmodes $\psi_{i}$ are ''nonlocal'' and extend over
several potential wells. As discussed above, we are interested in
calculating the number of atoms in each well as a function of
time. A more convenient basis in this case corresponds to a set of
modes $\phi_{i}$ localized in each potential well with the
corresponding destruction operators $b_{i}$ so that
\begin{equation}
    \hat \Psi = \sum_i \phi_i b_i.
\end{equation}
The operators $b$ are linear superpositions of the operators $a$,

\begin{equation}
    b_{i} = \sum_{j}u_{ji}a_{j},
\end{equation}
where $u$ is the transformation matrix determined by the condition
of localization of the modes $\phi_{i}$.

 Requiring that the destruction operators $b_{i}$ satisfy
the canonical commutation relations identical to those of
Eq.~(\ref{eqn:commutator_a})
\begin{eqnarray}
    &&[b_i, b_j^\dag] = \delta_{ij}, \nonumber \\
    &&[b_i, b_j] = 0, \label{eqn:commutator_b}
\end{eqnarray}
implies the unitarity of the transformation matrix $u$: $\sum_m
u_{mi} u_{mj}^* = \delta_{ij}$. For bound states all modes
$\psi_{i}$ can be chosen real and the transformation matrix $u$ can
be chosen real and orthogonal.

The transformation from the ''nonlocal'' basis $\psi_{i}$ to the
''local'' basis $\phi_{i}$ is given by the relations
\begin{equation}
    \phi_i = \sum_j u_{ji}^* \psi_j. \label{eqn:localmodes}
\end{equation}
The operators $b_{i}$ are associated with the local modes of the
$n$-well structure. For the purposes of the subsequent analysis we
will need to know only the lowest local mode in each potential well.
It means that there are $n$ local modes $\phi_{i}$ and the
coefficients $u_{ij}$ should be chosen so that the function
$\phi_{i}$ be localized in the $i$-th potential well.

\begin{figure}
\includegraphics[width=8.6cm]{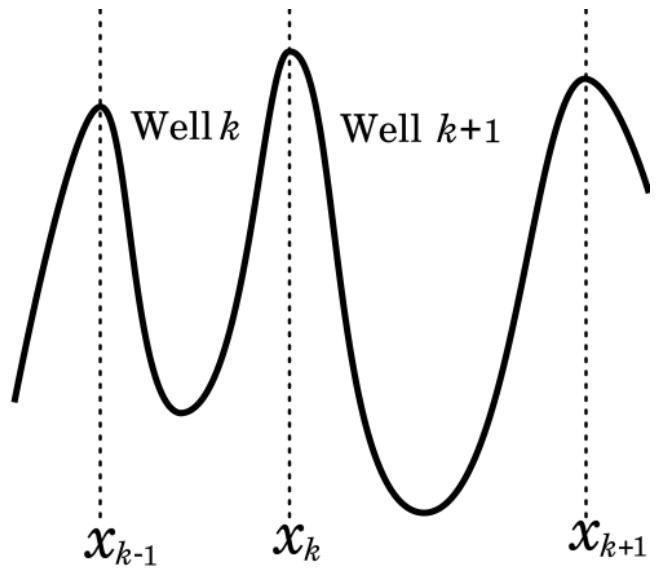}%
\caption{\label{fig:wells} A schematic of a multi-well non-symmetric
potential structure with two adjacent wells shown. The points
$x_{k-1}$, $x_{k}$ and $x_{k+1}$ are chosen between the wells where
the eigenmodes $\psi_{k}$ are exponentially small.}
\end{figure}
To quantify the degree of localization, we set points $x_0, x_1,
..., x_n$ somewhere between the wells where amplitudes of the modes
$\psi_k$ are exponentially small. This procedure is shown
schematically in Fig.~\ref{fig:wells}.  The degree of localization
of the mode $\phi_k$ in the $k$-th well is characterized by the
localization parameter
\begin{equation}
    f_k = \int_{x_{k-1}}^{x_k} dx |\phi_{k}|^2.
\end{equation}
The localization parameter $f_k$ can be rewritten in terms of the
transformation matrix $u$ as
\begin{equation}
    f_k = \sum_{m,l} u_{mk} u_{lk}^* J_k(l,m),
\end{equation}
where the matrix $J_k(l,m)$ is given by the expression
\begin{equation}
    J_k(l,m) = \int_{x_{k-1}}^{x_k} dx \psi_m^{*} \psi_{l}.
\end{equation}
To localize the modes $\phi_{k}$, we maximize the function
\begin{equation}
    f = \sum_k f_k = \sum_{kml} u_{mk} u_{lk}^* J_k(l,m)
\end{equation}
subject to constraints
\begin{equation}
    \sum_k u_{mk} u_{lk}^* = \delta_{ml}. \label{eqn:const1}
\end{equation}
The maximization results in the set of $n^{2}$ equations
\begin{equation}
    \sum_m \left[ J_j(i,m) - \lambda_{im} \right] u_{mj} = 0,
\label{eqn:min1}
\end{equation}
where $1 \le i,j \le n$ and where $\lambda_{im} = \lambda_{mi}^{*}$
are Lagrangian multipliers.

The set of equations (\ref{eqn:min1}) can be written in a more
transparent form as a set of $n$ matrix equations $(j = 1,2,
    \ldots n)$
\begin{equation}
    \left(\hat J_j -\hat \lambda \right) |u_j\rangle = 0,
\label{eqn:min2}
\end{equation}
where $\hat J_j$ and $\hat \lambda$ are Hermitian matrices with the
elements $J_j(i,m)$ and $\lambda_{im}$, respectively, and where
$|u_j\rangle$ is the column vector of $u$ with the elements $u_{mj}$
($m = 1, 2, \ldots n$). The equation of constraints, Eq.
(\ref{eqn:const1}), becomes
\begin{equation}
    \langle u_i | u_j \rangle = \delta_{ij}. \label{eqn:const2}
\end{equation}
In the limit of negligibly small coupling between the wells, the
column vectors $|u_{k}\rangle$ of the transformation matrix $u$ are
exact eigenvectors of the operators $\hat J_{j}$ because the latter
in this limit reduce to $\hat J_j = |u_j\rangle \langle u_j|$. The
matrix of Lagrange multipliers in this limit becomes the identity
matrix. This observation suggests that for nonzero coupling between
the wells the vectors $|u_{k}\rangle$ can be found perturbatively
starting from the eigenvectors $|w_{k}\rangle$ of $\hat J_{k}$ with
eigenvalues close to one:
\begin{equation}\label{eqn:eigenvectors_w}
    \hat J_{k}|w_{k}\rangle = \mu_{k}|w_{k}\rangle = (1 - \epsilon a_{k})
    |w_{k}\rangle,
\end{equation}
where $\epsilon \ll 1$ characterizes relative coupling strength
between the wells.
 The eigenvectors $|w_{k}\rangle$ form a nonorthogonal
basis set with $\langle w_{i}|w_{j}\rangle = O(\epsilon)$ for $i \ne
j$.

\begin{figure}
\includegraphics[width=8.6cm]{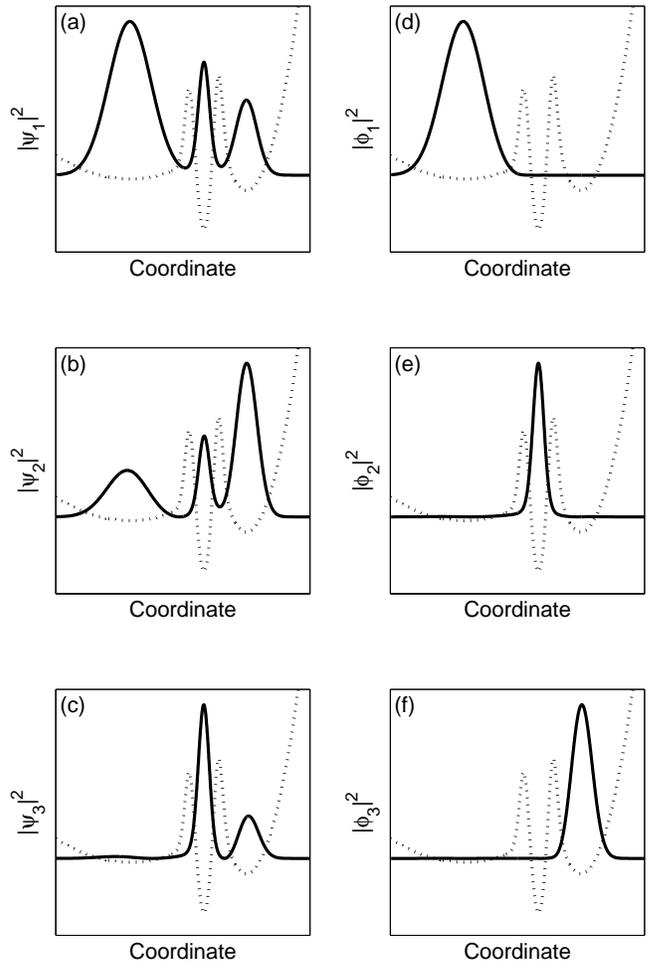}%
\caption{\label{fig:localmodes} An example of calculation of local
modes. Graphs (a)-(c) show the three lowest global eigenmodes of the
potential $V$ (dotted line). These eigenmodes are nonlocal with
large probability density in two or more wells. Graphs (d)-(f) show
local modes, which are linear combinations of the nonlocal
eigenmodes. }
\end{figure}

Solution of Eq.~(\ref{eqn:eigenvectors_w}) in the $|u_{k}\rangle$
basis to the first order in $\epsilon$ yields
\begin{equation}
    |w_{k}\rangle = |u_{k}\rangle + \sum_{j \ne k}|u_{j}\rangle
    \langle u_{j}|\hat J_{k}|u_{k}\rangle,
    \label{eqn:w_functionof_u}
\end{equation}
where we have used the fact that $\hat J_{k} - |u_{k}\rangle\langle
u_{k}| = O(\epsilon)$. Inversion of Eq.~(\ref{eqn:w_functionof_u})
yields
\begin{equation}
    |u_{i}\rangle = |w_{i}\rangle - \sum_{j \ne i}|w_{j}\rangle
    \langle u_{j}|\hat J_{i}|u_{i}\rangle,
    \label{eqn:u_functionof_w}
\end{equation}
Using the orthogonality conditions for $|u_{i}\rangle$ up to the
first order in $\epsilon$ and the condition $\langle u_{i}|\hat
J_{j}|u_{j}\rangle = \langle u_{i}|\hat J_{i}|u_{j}\rangle$ that
follows from Eq.~(\ref{eqn:min2}), results in the relation
\begin{equation}
    \langle u_{i}|\hat J_{j}|u_{j}\rangle = \frac{1}{2}\langle w_{i}|w_{j}\rangle
\end{equation}
yielding the final expression for the vectors $|u_{i}\rangle$ in
terms of $|w_{i}\rangle$:
\begin{equation}
    |u_i\rangle = |w_{i} \rangle - \frac{1} {2} \sum_{j \ne i} | w_{j}\rangle \langle w_{j}
    |w_{i}\rangle. \label{eqn:trans_final}
\end{equation}
To calculate the local modes, one thus finds eigenvectors
$|w_{i}\rangle$ of $\hat J_i$ with eigenvalues close to one for $i =
1,2,\cdots n$.  The columns of the transformation matrix are then
given by Eq.~(\ref{eqn:trans_final}). The local modes are found
using Eq.~(\ref{eqn:localmodes}). An example of such calculation is
shown in Fig.~\ref{fig:localmodes}.
%

Overall signs of the local modes $\phi_{i}$ in
Eq.~(\ref{eqn:localmodes}) are arbitrary being determined by sign
choices for the global modes $\psi_{i}$. These signs can be changed
if needed because if $\phi_{i}$ is a local eigenmode, so is
$-\phi_{i}$. Changing the sign of $\phi_{i}$ amounts to changing the
sign of the $i$-th row of the transformation matrix $u$ which leaves
it unitary. To be able to unambiguously specify the value of the
relative phase shift between condensates in different potential
wells, the overall signs of the local modes $\phi_{i}$ will be fixed
by requiring that each eigenmode $\phi_{i}$ be positive in the
region of its localization between $x_{i}$ and $x_{i+1}$.

In terms of the destruction operators $b_{i}$ of the local modes the
Hamiltonian Eq.~(\ref{eqn:SQ_H1}) can be written as
\begin{equation}
H = \sum_{ijk} \hbar \Omega_{k} u_{ki} u_{kj}^{*} b_{i}^\dag b_{j} +
\frac {U_0}{2} \sum_i \kappa_{i}\left(b_i^\dag\right)^{2} b_{i}^{2},
\label{eqn:SQ_H3}
\end{equation}
where $\Omega_i$ is the eigenfrequency of the $i$-th mode $\psi_{i}$
given by Eq.~(\ref{eqn:SQ_H2}), $u_{ij}$ is the transformation
matrix and $\kappa_i$ is the overlap integral
\begin{equation}
    \kappa_i = \int dx |\phi_i|^4. \label{eqn:SQ_kappa}
\end{equation}
The equations of motion for the operators $b_i$ in the Heisenberg
picture are given by
\begin{equation}\label{eqn:Heisenberg_eqs_motion}
    i \hbar \frac{d}{dt} b_i = \sum_{jk} \hbar \Omega_k u_{ki} u_{kj}^* b_j +
    U_0 \kappa_i b_i^\dag b_{i}^{2}.
\end{equation}
The diagonal terms
\begin{equation}\label{eqn:SQ_lambda}
    \Lambda_{i} = \sum_{k}\Omega_{k}|u_{ki}|^{2}
\end{equation}
in Eq.~(\ref{eqn:SQ_H3}) have the meaning of eigenfrequencies of the
local eigenmodes in the absence of coupling between the wells and
the nondiagonal terms
\begin{equation}\label{eqn:SQ_Delta}
    \Delta_{ij} = \sum_{k}\Omega_{k}u_{ki}u_{kj}^{*}
\end{equation}
are coupling frequencies between the $i$-th and $j$-th wells. Since
for bound states $u$ can be chosen real, the matrix of the coupling
frequencies is real and symmetric: $\Delta_{ij} = \Delta_{ji}$. The
coupling strength is exponentially dependent on the distance between
the wells and usually only the nearest-neighbor coupling should be
taken into account.

\section{Three-well structure}\label{sec:3well_structure}

In the following we shall specialize our analysis to the case of a
potential consisting of three potential wells. These will be
referred to as the left, middle and right well, respectively. The
left well serves as a source of atoms. The number of atoms $N_{l}$
in this well is kept nearly constant and is considerably larger than
the number of atoms initially placed and subsequently tunneling into
the middle or the right wells. The dynamics in the left well is
therefore unaffected by that in the other two wells. This dynamics
is factored out and the destruction operator for the left well
$b_{l}$ is replaced by a c-number: $b_l = \rightarrow \sqrt{N_l}$.

The Hamiltonian Eq.~(\ref{eqn:SQ_H3}) reduces to
\begin{eqnarray}\label{eqn:SQ_H4}
    &&H = \hbar (\Lambda_{m} - \mu) b_{m}^\dag b_{m}
    + \hbar (\Lambda_{r} - \mu) b_{r}^\dag b_{r} \nonumber \\
    &&+ \hbar (\Delta_{lm}\sqrt{N_{l}} b_{m} + \Delta_{mr} b_{m}^\dag b_{r} +
    h.c.) \nonumber \\
    &&+ \frac{U_{0}}{2} \kappa_{m} \left(b_{m}^\dag\right)^{2} b_{m}^{2} +
    \frac{U_{0}}{2} \kappa_{r} \left(b_{r}^\dag\right)^{2} b_{r}^{2}
\end{eqnarray}
where $\Lambda_{i}$, $\Delta_{i}$ and $\kappa_{i}$ are given by
Eqs.~(\ref{eqn:SQ_lambda}), (\ref{eqn:SQ_Delta}) and
(\ref{eqn:SQ_kappa}), respectively, $h.c.$ means Hermitian conjugate
and $\mu = \hbar \Lambda_{l} + \kappa_{l}U_{0}N_{l}$.

As discussed at the end of Sec.~\ref{sec:eqns_of_motion}, the
overall sign of the local modes $\phi_{i}$ has been fixed by
requiring that they be positive in the region of their localization.
With this choice, the coupling frequencies $\Delta_{ij}$ between
different wells (see Eq.~(\ref{eqn:SQ_Delta})) are negative. This is
easily ascertained using the simplest example of a symmetric
two-well structure where the two local modes are proportional to a
sum and a difference of the two global modes. Normalizing the
Hamiltonian Eq.~(\ref{eqn:SQ_H4}) to the positive energy $-\hbar
\Delta_{mr} = \hbar |\Delta_{mr}|$ brings it to its final
dimensionless form

\begin{eqnarray}\label{eqn:SQ_dimensionless_H}
    \frac{H}{\hbar |\Delta_{mr}|} &=&  \omega_m b_m^\dag b_m + \omega_r b_r^\dag
    b_r -
    (D b_m +  b_m^\dag b_r + h.c.) \nonumber \\
    &+& \  \frac{Z_m}{2} \left(b_m^\dag\right)^{2}b_{m}^{2} +
    \frac{Z_r}{2} \left(b_r^\dag\right)^{2}b_{r}^{2}, \label{eqn:H_dim_less}
\end{eqnarray}

where $Z_i = -U_0 \kappa_i /\hbar \Delta_{mr}$, $\omega_i = (\mu -
\Lambda_i)/ \Delta_{mr}$ and $D = \Delta_{lm} \sqrt{N_l}/
\Delta_{mr}$.

The Heisenberg equations of motion for the destruction operators
$b_{l}$ and $b_{r}$ (\ref{eqn:Heisenberg_eqs_motion}) in the
dimensionless variables take the form
\begin{eqnarray}\label{eqn:SQ_bMbL_equations}
    i \frac{d }{ d\tau} b_m &=&  (\omega_m + Z_m b_m^\dag b_m) b_m - D - b_r \nonumber \\
    i \frac{d }{ d\tau} b_r &=&  (\omega_r + Z_r b_r^\dag b_r) b_r - b_m,
\end{eqnarray}
where the dimensionless time $\tau$ is given by the relation $\tau =
|\Delta_{mr}|t$.

\subsection{Mean-field}\label{subsec:meanfild}

In this section we shall present results of analysis of
Eq.~(\ref{eqn:SQ_bMbL_equations}) in the mean-field limit
corresponding to relatively large atomic populations in all wells,
when the operators $b_{m}$ and $b_{r}$ can be treated as complex
numbers.

\begin{figure}
\includegraphics[width=8.6cm]{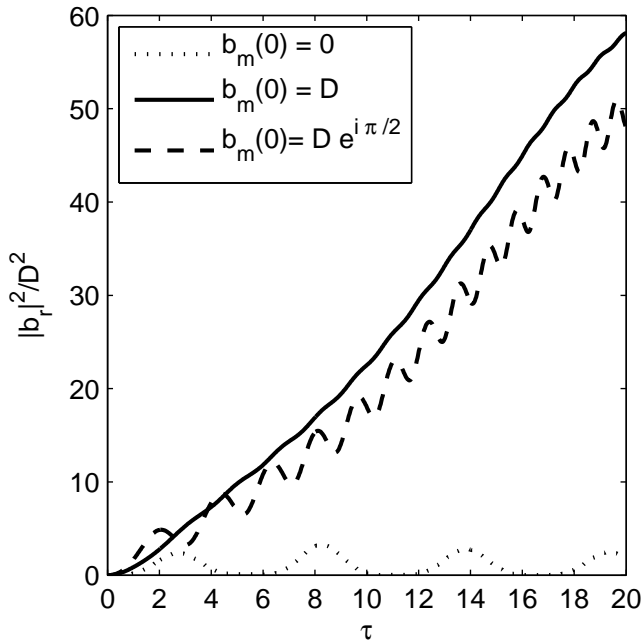}%
\caption{\label{fig:P_of_time} The number of atoms in the right well
as a function of interaction time for different initial number of
atoms in the middle well.  The dotted curve corresponds to initially
empty middle well, $b_m(0) =0$, and the solid curve to $b_{m}(0) =
D$. The dashed curve corresponds to the initial condition $b_{m}(0)
= D\exp(i\pi/2)$. For all curves $\omega_m = -1.3$, $\omega_{r} =
0.5$, $Z_m D^2 = 1$ and $Z_{r} D^{2} = 0$.}
\end{figure}

Figure~\ref{fig:P_of_time} demonstrates control of atomic population
in the right well by population in the middle well with the absolute
gain that is considerably larger than one. Parameters for
Fig.~\ref{fig:P_of_time} are $\omega_m = -1.3$, $\omega_{r} = 0.5$,
$Z_m D^2 = 1$ and $Z_{r} D^{2} = 0$. The right well is initially
empty, $b_{r}(0) = 0$. Parameters of the wells are chosen so that if
no atoms are initially placed in the middle well ($b_{m}(0) = 0$),
the tunneling from the source (the left well) to the middle well is
strongly suppressed and the population in the right well remains at
a low level. This situation is illustrated by a dotted line in
Fig.~\ref{fig:P_of_time}.

Placing some number of atoms in the middle well results in a much
larger tunneling rate from the left to the right well through the
middle well as shown by a solid line corresponding to the initial
condition $b_{m}(0) = D$. The increase in the tunneling rate can be
observed for a range of values of the relative phase of the
condensates in the left and middle wells. The dashed curve obtained
for the initial condition $b_{m}(0) = D\exp(i\pi/2)$, i.e.,
corresponding to the $\pi/2$ relative phase shift between the
condensates in the left and middle wells, exhibits qualitatively
similar behavior. Note that the output number of atoms in the right
well ($\tau = 20$) is about $50-60$ times larger than the input
number of atoms in the middle well. In other words, the output
number of atoms in the right is controlled by that in the middle
well with the absolute gain $G = N_{r,out}/N_{l,in} \approx 50-60$.

\begin{figure}
\includegraphics[width=8.6cm]{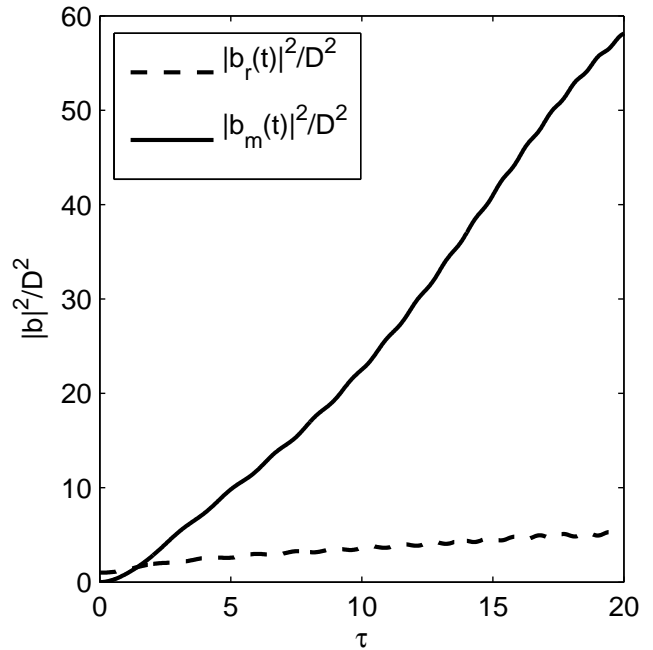}%
\caption{\label{fig:r_m_of_time} The number of atoms in the middle
(dashed) and right (solid curve) well as a function of interaction
time for $\omega_m = -1.3$, $\omega_{r} = 0.5$, $Z_m D^2 = 1$,
$Z_{r} D^{2} = 0$ and $b_{m} = D$}
\end{figure}

Populations in the middle and right wells as functions of the
interaction time are shown if Fig.~\ref{fig:r_m_of_time} for
$b_{m}(0) = D$. All other parameters are the same as in previous
graphs. The solid curve is the population of the right well and the
dashed curve is the population of the middle well.  Figure
\ref{fig:r_m_of_time} demonstrates that the population of the middle
well stays about an order of magnitude below that for the right
well. The middle well serves a gate controlling the rate of atomic
flow from the source to the right well. The atoms tunneling from the
source to the right well pass through the middle well without being
accumulated there.

\begin{figure}
\includegraphics[width=8.6cm]{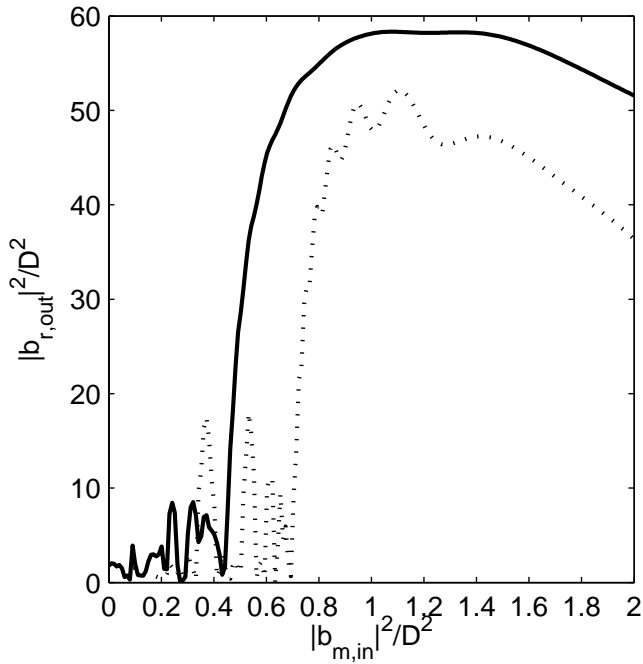}%
\caption{\label{fig:growth_number} The output number of atoms in the
right well ($\tau = 20$) as a function of the number of atoms
initially placed in the middle well.}
\end{figure}

The output number of atoms in the right well as a function of the
input number of atoms in the middle well is shown in Fig.
\ref{fig:growth_number}. Parameters for this figure are the same as
for Fig.~\ref{fig:P_of_time}, i.e., $\omega_m = -1.3$, $\omega_r =
0.5$, $Z_m D^2 = 1$ and $Z_r D^2 = 0$. The solid curve corresponds
to the zero initial phase shift between the condensates in the left
and middle wells and the dotted curve to the $\pi/2$ shift. This
figure demonstrates rapid switching from small to large tunneling
rates in the region around $|b_{m,in}|^{2}D^{2} \approx 0.5$ with
subsequent saturation at the level $G = N_{r,out}/N_{m,in} \approx
50-60$. In the switching region, a small change in the population of
the atoms in the middle well results in a large difference in the
population in the right well.

\begin{figure}
\includegraphics[width=8.6cm]{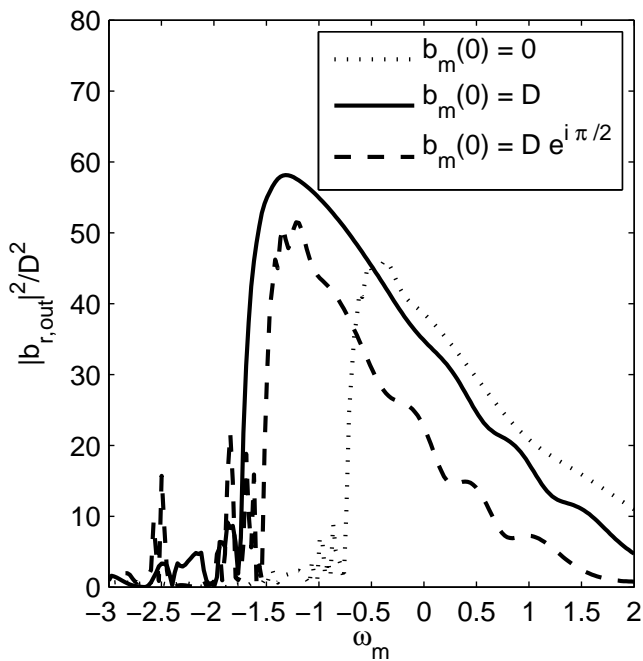}%
\caption{\label{fig:P_of_W_m}  The output number of atoms in the
right well ($\tau = 20$) as a function of the frequency of the
middle well $\omega_{m}$.  The dotted line corresponds to the
initially empty middle well, $b_{m}(0) = 0$, the solid line to $b_m
= D$, and the dashed line to $b_m = D\exp(i\pi/2)$. }
\end{figure}

Figure \ref{fig:P_of_W_m} shows output population in the right well
($\tau = 20$) as a function of the modal frequency $\omega_{m}$ of
the middle guide for different values of the input number of atoms
in the middle well. This figure demonstrates switching for different
values of the number of atoms initially in the middle well. The
dotted line corresponds to initially empty middle well. For this
curve, the maximum tunneling rate corresponds to the region around
$\omega_m = -0.5$. If the frequency of the middle well is lowered
beyond this value, the number of atoms that tunnel into the right
well becomes small. The solid curve corresponds to the initial
condition $b_m(0) = D$. This curve is qualitatively similar to that
for an initially empty middle well, but the maximum has moved to a
lower value of $\omega_m = -1.3$. The dashed curve corresponds to
the initial condition $b_m(0) = D\exp(i\pi/2)$. This curve still has
a maximum around $\omega_m = -1.3$. For this value of $\omega_m$,
the number of atoms that tunnel into the right well when the initial
atoms have either zero or $\pi/2$ phase shift is about the same. For
the initially empty middle well, the population in the right well
remains small.

\begin{figure}
\includegraphics[width=8.6cm]{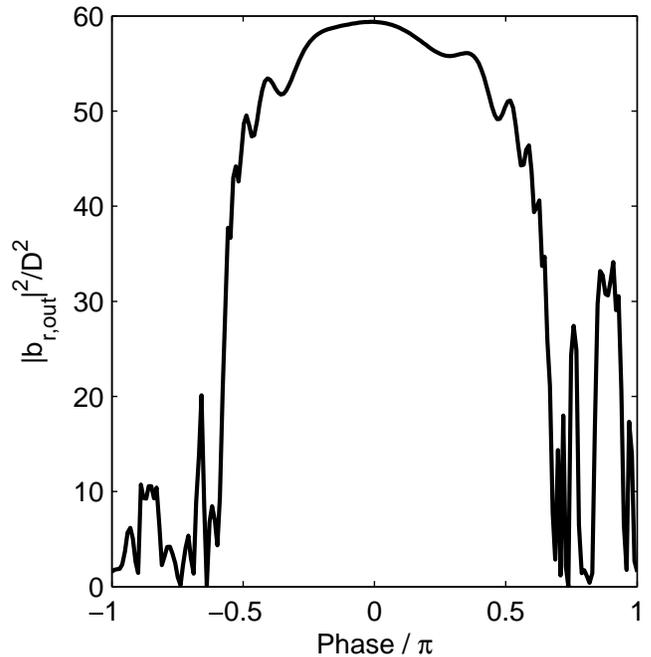}%
\caption{The output number of atoms in the right well as a function of the relative phase of the atoms placed in the middle well.\label{fig:growth_phase}}
\end{figure}

As opposed to an electronic transistor, the amplification and
switching in the three-well structure is a coherent effect and
depends on the relative phase between the condensates in the left
and middle wells. To investigate sensitivity of the previously
obtained results to the value of the relative phase angle, we kept
the input number of atoms in the middle well fixed at $|A_m(0)/D|^2
= 1.2$ and changed the relative phase angle. The results are given
by Fig.~\ref{fig:growth_phase} showing the output number of atoms in
the right well ($\tau = 20$) as a function of the phase angle. All
other parameters are the same as in previous figures. Figure
\ref{fig:growth_phase} demonstrates that in the amplification regime
the number of atoms that tunnel into the right well is nearly
independent of the initial phase angle, as long as this angle is
roughly in the range between $-\pi/2$ and $\pi/2$.

\subsection{Second-quantization results}\label{subsec:SQ}

This section presents results of analysis of the three-well
structure in the framework of the second-quantization formalism.
This allows us to estimate the region of applicability of the
mean-field approach of section \ref{subsec:meanfild}, evaluate
intrinsic quantum-mechanical uncertainty due to finite number of
atoms and extend previous results to the limit of small number of
atoms.

In the dimensionless variables, the state vector of the system
$|\psi(t)\rangle$ evolves according to the equation
\begin{equation}
    i \frac {d}{d \tau}|\psi\rangle = H|\psi\rangle, \label{eqn:S_eq}
\end{equation}
where $H$ is the second-quantized Hamiltonian given by
Eq.~(\ref{eqn:H_dim_less}). The state vector can be represented in
terms of the joint number states $|n_m, n_r \rangle$ as
\begin{equation}
    |\psi\rangle = \sum_{i,j} c_{i,j} |n_i, n_j\rangle,
\end{equation}
with the decomposition coefficients given by $c_{m,r} = \langle n_m,
n_r | \psi \rangle$. Equation (\ref{eqn:S_eq}) is transformed to the
set of ordinary differential equations that describe the evolution
of the decomposition coefficients
\begin{equation}
    i\frac{d}{d\tau}c_{i,j} = \sum_{i,j,k,l}\langle n_i,n_j|H
    |n_k,n_l \rangle c_{k,l}. \label{eqn:ode1}
\end{equation}

In simulations, the set of Eqs.~(\ref{eqn:ode1}) has been truncated
by keeping only the values of $n_{m}$ and $n_{r}$ such that $n_r +
n_m \le N_{max}$. The value of $N_{max}$ was chosen so that
$N_{max}$ was several times larger than  the sum $\langle n_r
\rangle + \langle n_m \rangle$.

Initial conditions for the system of equations Eq. (\ref{eqn:ode1})
corresponded to zero initial number of atoms in the right well with
the atoms in the middle well being in a coherent state:
\begin{equation}
    |\psi(0) \rangle = e^{-|\alpha|^2} \sum_{n = 0}^{N_{max}}
    \frac{\alpha^{n}} {\sqrt{n!}}|n,0\rangle.
\end{equation}
Here the complex parameter $\alpha$ is given by $\alpha =
\sqrt{\langle N_m\rangle(0)} e^{i\varphi}$, where $\langle
N_m\rangle(0)$ is the average number of atoms initially placed in
the middle well and $\varphi$ is the phase difference between the
atoms in the middle and left wells.

\begin{figure}
\includegraphics[width=8.6cm]{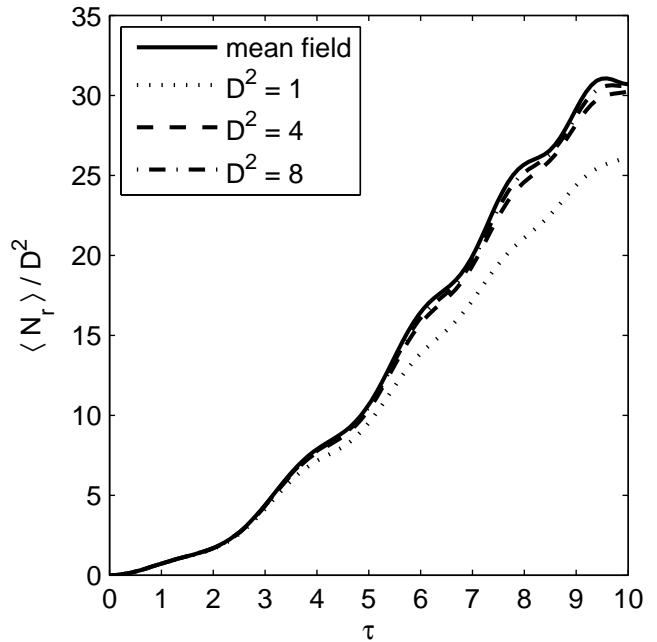}%
\caption{\label{fig:SQ_A_of_T} The average number of atoms in the
right well as a function of the interaction time for three different
values of the coupling between the left and the middle wells. Atoms
in the middle well are initially in a coherent state with $\langle
N_{m} \rangle (0) = D^2$ and zero phase. The solid line is the
result of the mean-field calculation, the dotted line is the result
of the second-quantization calculation with $D^2 = 1$, the dashed
line corresponds to $D^2 = 4$, and the dash-dotted one to $D^2 =
8$.}
\end{figure}

The transition to the mean-field limit corresponds to increasing the
input number of atoms in the middle well $\langle N_{m} \rangle (0)$
while keeping the ratio $\omega_{m}/Z_{m}\langle N_{m}\rangle (0)$
constant. Equations (\ref{eqn:H_dim_less}) show that the results of
action of the destruction operators on the state vector scales as
$D$ provided the parameters $Z_{m}D^{2}$ and $Z_{r}D^{2}$ are kept
constant. Thus, the transition to the mean-field limit can be
implemented by setting the initial number of atoms in the middle
well proportional to $D^{2}$ and increasing the value of coupling
$D$ between the left and middle wells while keeping the parameters
$Z_{m}D^{2}$ and $Z_{r}D^{2}$ constant.

The average number of atoms in the right well $\langle N_{r}
\rangle$ as a function of the interaction time is shown in
Fig.~\ref{fig:SQ_A_of_T} for three different values of $D^2$.  The
parameters for this figure are $\omega_r = 1$, $\omega_m=-0.5$, $Z_m
D^2=1/4$, $Z_{r}D^{2} = 0$ and $|\alpha|^{2} = \langle N_{m} \rangle
(0) = D^2$. The phase angle of the coherent state is zero. The solid
line is the mean-field limit. The dotted, dash-dotted and dashed
lines correspond to $D^2 = 1$, $D^2 = 4$ and $D^2 = 8$,
respectively. Figure \ref{fig:SQ_A_of_T} demonstrates good
convergence of the second-quantization results to the mean field
limit as $D^{2}$ is increased. The $D^{2} = 1$ curve deviates from
the mean-field limit for large value of $\tau$, but all other curves
lie progressively closer to the mean-field curve as the parameter
$D$ increases.

\begin{figure}
\includegraphics[width=8.6cm]{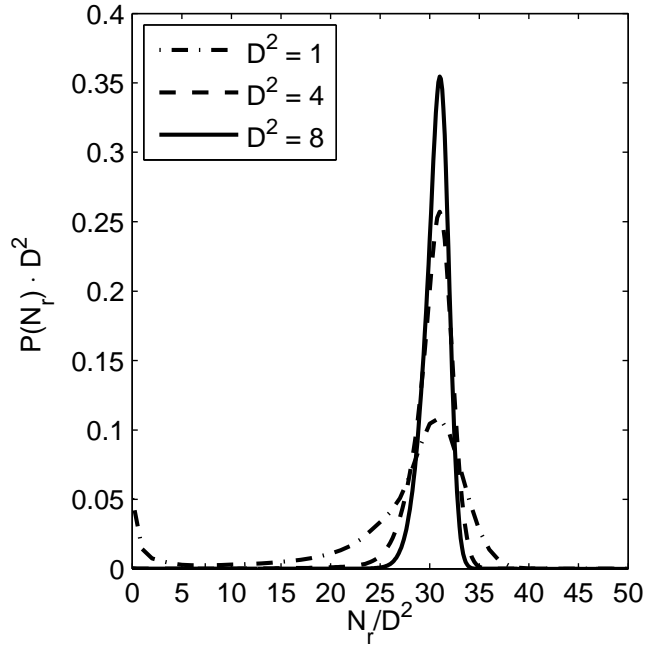}%
\caption{\label{fig:SQ_P_of_N}  The probability $P(N_r)$ of finding
$N_{r}$ atoms in the right well ($\tau = 10$). Atoms in the middle
well are initially in a coherent state with $\alpha = D$. The
dash-dotted line corresponds to $D^2 = 1$, the dashed line to $D^2 =
4$ and the solid line to $D^2 = 8$.}
\end{figure}
The output ($\tau = 10$) probability distribution $P(N_r)$ of
finding $N_r$ atoms in the right well versus $N_r$ is shown if
Fig.~\ref{fig:SQ_P_of_N}.  The dash-dotted line corresponds to
$D^2 = 1$, the dashed line to $D^2= 4$ and the sold line to $D^2 =
8$. Since the number of output atoms scales as $D^{2}$, the
horizontal axes is scaled as $N_{r}/D^{2}$ to keep position of the
maximum and the width of the curves more of less the same for
different values of $D^{2}$. As a result, the vertical axes shows
not $P(N_{r})$, but the product $P(N_{r})D^{2}$ to keep the height
of the curves approximately the same for different values of
$D^{2}$. The total ''area under the curve'' (strictly speaking it
is a sum, not an integral) for all curves is equal to one. The
dash-dotted curve corresponding to $D^2 = 1$ shows bimodal
distribution with a relatively large probability of finding atoms
near zero $N_{r}$ in addition to the main peak near $N_{r}/D^{2}
\approx 31$, the latter being very close to the mean-field result.
The difference between the mean-field and second-quantization
results, previously seen in Fig.~\ref{fig:SQ_A_of_T} for $D^{2} =
1$, is due to the part of the probability distribution near zero
that pulls down the average. As the coupling $D$ to the source is
increased, only the single-humped part of the probability
$P(N_{r})$ centered at the mean-field result remains. The output
relative standard deviation $\Delta N_{r}/\langle N_{r} \rangle$
is equal to $0.35$, $0.08$ and $0.04$ for $D^{2} = 1, 4$ and $8$,
respectively.

Comparison of the mean-field and second-quantization results carried
out for the same parameters as above but the relative phase angle
between the condensates equal to $\phi = \pi/2$ yielded conclusions
very similar to those summarized by Figs.~\ref{fig:SQ_A_of_T} and
\ref{fig:SQ_P_of_N}.

\begin{figure}
\includegraphics[width=8.6cm]{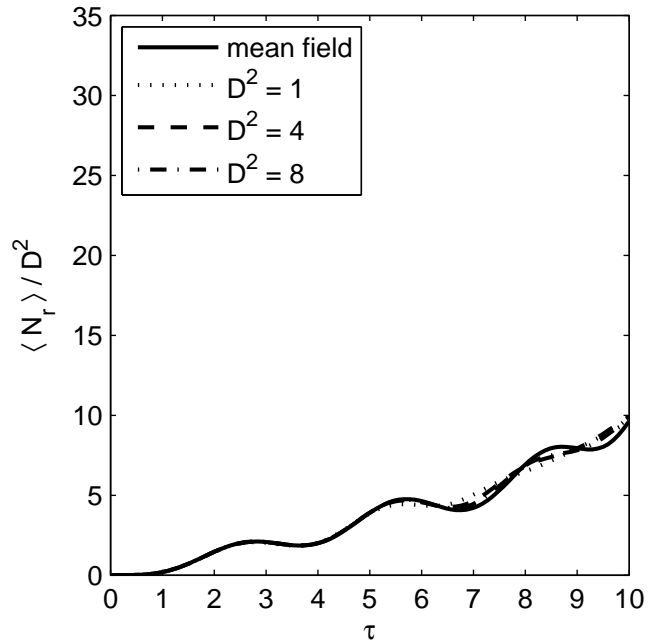}%
\caption{\label{fig:eSQ_A_of_T} The average number of atoms in the
right well as a function of the interaction time for three different
values of the coupling between the left and the middle wells. The
middle well is initially empty. The solid curve is the result of the
mean-field calculation, the dotted curve is the result of the
second-quantization calculation with $D^2 = 1$, the dashed curve
corresponds to $D^2 = 4$, and the dash-dotted curve to $D^2 = 8$.}
\end{figure}

\begin{figure}
\includegraphics[width=8.6cm]{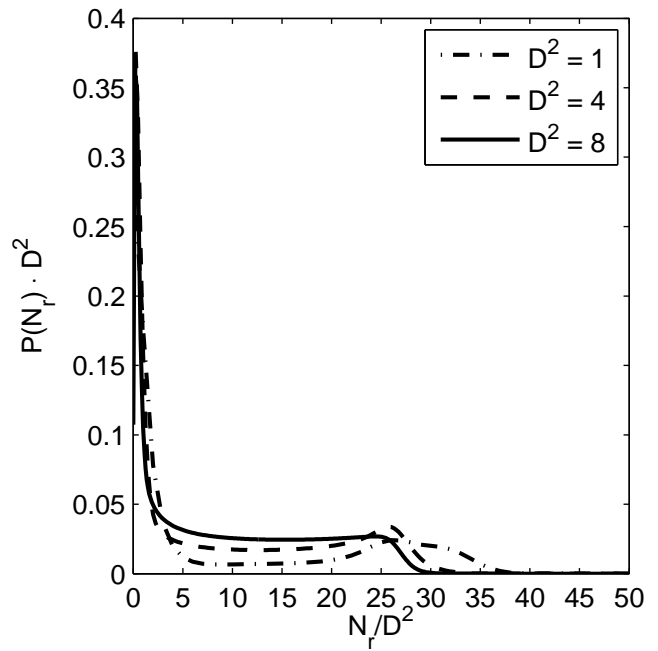}%
\caption{\label{fig:eSQ_P_of_N} The probability $P(N_r)$ of finding
$N_{r}$ atoms in the right well ($\tau = 10$). The middle well is
initially empty. The dash-dotted line corresponds to $D^2 = 1$, the
dashed line to $D^2 = 4$ and the solid line to $D^2 = 8$.}
\end{figure}

Figures \ref{fig:eSQ_A_of_T} and \ref{fig:eSQ_P_of_N} parallel
analysis of Figs.~\ref{fig:SQ_A_of_T} and \ref{fig:SQ_P_of_N} for
the case when the middle well is initially empty, $|\alpha|^{2} =
\langle N_{m} \rangle (0) = 0$. These figures are aimed at verifying
that the rapid switching from low to high-amplification regime
predicted by the theory in the mean-field limit can be also realized
with only few controlling atoms.

Figure \ref{fig:eSQ_A_of_T} demonstrates convergence of the
second-quantization results to the mean-field limit for $\langle
N_{m} \rangle (0) = 0$ as $D^{2}$ is increased. This convergence is
similar to that shown in Fig.~\ref{fig:SQ_A_of_T} except in this
case the second-quantization approach gives values somewhat larger
than the mean-field limit. The reason is explained by
Fig.~\ref{fig:eSQ_P_of_N}, which shows the probability $P(N_r)$ of
finding $N_r$ atoms in the right well at $\tau = 10$. The
probability $P(N_{r})$ has a pronounced spike at low values of
$N_{r}$. Another noticeable feature of Fig.~\ref{fig:eSQ_P_of_N} is
a wide, nearly flat pedestal extending from small values of $N_{r}$
to a maximum value that is about twice larger than the average (cf.
Fig.~\ref{fig:eSQ_A_of_T}). The maximum value slightly decreases as
$D$ increases. This explains why the second-quantization results are
larger than the mean-field results. The one-humped shape of
$P(N_{r})$ in Fig.~\ref{fig:SQ_P_of_N} means that the uncertainty in
the output number of atoms in the high-amplification regime is small
for even a few controlling atoms in the middle well. Figures
\ref{fig:eSQ_A_of_T} and \ref{fig:eSQ_P_of_N} show that the
low-amplification region is characterized by both low average number
of output atoms and by large uncertainty corresponding to the
average. Indeed, the output relative standard deviation $\Delta
N_{r}/\langle N_{r} \rangle$ for the results of
Fig.~\ref{fig:eSQ_P_of_N} is equal to $1.3$, $1.0$ and $0.9$ for
$D^{2} = 1, 4$ and $8$, respectively. These results are in contrast
to those for the large-amplification regime of
Fig.~\ref{fig:SQ_P_of_N}, where the standard deviation rapidly goes
down as the parameter $D^{2}$ increases.

\section{Discussion}\label{sec:discussion}

The analysis of Sec.~\ref{sec:3well_structure} demonstrates that a
Bose Einstein condensate in a three well potential shows
transistor-like behavior with the middle well acting as a gate
controlling the flux of atoms from the source to the drain. In this
section we present estimates of the characteristic tunneling time
for a trapped atom transistor and discuss possible gain in the total
number of atoms. The analysis will be extended to the case of a
waveguide device, where estimates will be presented for the
tunneling time, the length of the device and the gain in the output
flux of atoms. Finally, we summarize the results obtained.

\subsection{Trapped atom transistor}

The parameter $D$ in Eq.~(\ref{eqn:SQ_bMbL_equations})
characterizes the strength of coupling of the source (left well)
to the gate (medium well). It is reasonable to expect that the
operational parameters of the BEC transistor are such that the
contributions of the nonlinear and linear terms in
Eq.~(\ref{eqn:SQ_bMbL_equations}) are of the same order of
magnitude, i.e. $Z_{m}D^{2} \approx 1$.

The growth curve, shown if Fig. \ref{fig:growth_number} shows the
final population of the right well as a function of the initial
population of the middle well, where the atoms are held in the
traps for a dimensionless time $\tau = 20$.
This figure demonstrates that a change in the population of the
middle well from $N_{m} = 0.4 D^2$ to $N_{m} = 0.8 D^2$ results in
a change in the final population of the right well from $N_{r}
\approx 10 D^2$ to $N_{r} \approx 60 D^2$.  The maximum number of
atoms that tunnel into the right occurs when the  number of atoms
initially in the middle well is $N_m \approx D^2$. We will refer
to this number as the saturation number. For example, if we take
$D^2 = 10$, a change from $4$ to $8$ atoms in the middle well
results in a change from $100$ to $600$ atoms in the right well.

Assume that the potential energy of the middle well is a cigar
shaped potential of the form
\begin{equation}
V(r_{\perp}, z) = \frac {1} {2} m \left( \omega_{\perp}^{2} r_{\perp}^{2} + \omega_{z}^{2} z^{2} \right),
\end{equation}
where $r_{\perp}$ is the coordinate in the radial direction, and
$z$ is the coordinate in the axial direction.  For this potential,
the overlap integral given by the Eq.~(\ref{eqn:SQ_kappa}) can be
evaluated as
\begin{equation}
\kappa_{m} = \frac{1}{(2 \pi)^{3/2}}\frac{1}{a_{\perp}^{2} a_{z} },
\label{eqn:overlap1}
\end{equation}
where $a_{\perp} = \sqrt{\hbar / m \omega_{\perp}}$ and $a_{z} =
\sqrt{\hbar / m \omega_{z}}$ are the harmonic oscillator lengths.
The nonlinearity parameter $Z_{m}$ in
Eq.~(\ref{eqn:SQ_dimensionless_H}) is given by the expression
\begin{equation}
Z_{m} = \frac {U_{0} \kappa_{m} } {\hbar |\Delta_{mr}|},
\label{eqn:Zm},
\end{equation}
where $U_{0} = 4 \pi a_s \hbar^{2}/ m$.
Using Eq. (\ref{eqn:overlap1}) in Eq. (\ref{eqn:Zm}) allows one to
express the tunneling frequency between the middle and the right
wells as
\begin{equation}
|\Delta_{mr}| = {a_{s}}\frac  {N_{m}}{a_{z}}  \omega_{\perp},
\label{eqn:D_mr}
\end{equation}
where we have also used $Z_{m} N_{m} \approx 1$ to eliminate $Z_{m}$ in favor of $N_{m}$.

If the middle well is a spherical trap with $\omega_z =
\omega_{\perp} = 2 \pi \times 10^3 Hz$ and the saturation number is
$D^2 = 10$ the tunneling frequency between the middle and right well
is
\begin{equation}
\Delta_{mr} \approx \pi \times 10^2 rad / sec.
\end{equation}
The dimensional time that it takes for atoms to tunnel from the left to the right wells is
\begin{equation}
 t  \approx 2 \times 10^{-1} sec.
\end{equation}

In other words, for the parameters chosen a trapped atom
transistor can distinguish between 4 and 8 atoms in the gate with
the characteristic operational time of $10^{-1} sec$. This time
can be decreased either by increasing the frequency of the trap or
increasing the value of the saturation number.

\subsection{Waveguide transistor}

In a waveguide transistor the potential wells of Fig.~\ref{fig:toon}
are the three guides that run parallel to each other for the
distance $L$. The interaction time $T = L/v$ is determined by the
speed of flow $v$ of the BEC in the guides. The field operator for
this configuration can be expressed as
\begin{equation}
\hat \Psi(\textbf{r},t) = \exp(i k_p z - i \omega_p t) \hat \psi(\textbf{r},t),
\end{equation}
where $k_p$ and $\omega_p = \hbar k_p^2 / 2m$ and the carrier wave
number and frequency, respectively, and $\hat \psi$ is the
field-operator envelope.

The Heisenberg equations of motion for the field operator $\hat
\psi$ in the co-propagating frame $t' = t$, $z' = z - vt$ is of the
form
\begin{equation}
i \hbar \frac {\partial} {\partial t} \hat \psi =
\left[ - \frac {\hbar^2 }{2 m} \left( \nabla_\perp^2 + \frac {\partial^2} {\partial z^2}  \right)
+ V(\textbf{r}_\perp) + U_0 \hat \psi^{\dag} \hat \psi\right]\hat \psi, \label{eqn:Heisenberg_FO}
\end{equation}
where $v = \hbar k_p / m$ is the velocity of the condensate and the primes have been omitted.

Changes in density as the condensate propagates through the
transistor occur at a length scale $L_{BEC}$. We assume that the
kinetic energy associated with the longitudinal direction is small
in comparison with the characteristic energy $\hbar \Omega$
associated with the transverse eigenmodes of the transistor
\begin{equation}
\hbar \Omega \gg \frac {\hbar^2} {2 m L_{BEC}^2}. \label{eqn:ineq1}
\end{equation}
Next, we require that $L_{BEC}$ does not change appreciably during
the time interval $L/v$ that it takes the condensate to propagate
through the transistor,
\begin{equation}
\left| \frac { \partial }  {\partial t} \ln {L_{BEC} }\right| \ll \frac {v} {L} \label{eqn:ineq2}
\end{equation}

With Eqs. (\ref{eqn:ineq1}) and (\ref{eqn:ineq2}) fulfilled, the
dispersive term ($\partial^2 / \partial z^2$) in
Eq.~(\ref{eqn:Heisenberg_FO}) can be neglected and the coordinate
$z$ becomes a parameter. Propagation of different "slices" of the
condensate (parametrized by the coordinate $z$) through the
transistor can be analyzed independently.

Represent the field operator $\hat \psi$ as
\begin{equation}
\hat \psi = \sum_{i}  \phi_i(\textbf{r}_\perp) b_i(z,t),
\label{eqn:field_op_decomp}
\end{equation}
where $\phi(\textbf{r}_\perp)$ is the $i$-th transverse local mode
and $b_i(z,t)$ is the destruction operator that destroys an atom
in the $i$-th local mode at the coordinate $z$.  Note that $b_i$
now has dimension of $m^{-1/2}$.

Using Eq.~(\ref{eqn:field_op_decomp}) in Eq.
(\ref{eqn:Heisenberg_FO}) with the dispersive term dropped,
results in the equations of motion in the Heisenberg picture for
the operators $b_i$ that are of the same form as Eq.
(\ref{eqn:Heisenberg_eqs_motion}):
\begin{equation}
i \hbar \frac {d } {dt} b_i(z,t)  = \sum_{i,j} \hbar \Omega_k u_{ki}u_{kj}^* b_j + U_0 \kappa_{i} b_{i}^\dag b_{i},
\end{equation}

As in the case of a trapped device, the left guide will be treated
as a reservoir of atoms corresponding to the replacement $b_l
\rightarrow \sqrt{n_l}$, where $n_l$ is the density of atoms (number
of atoms per unit length). The equations of motion for the atoms in
the middle and right guide, in dimensionless form, become
\begin{eqnarray}
    i \frac{d }{ d\tau} b_m(z,t) &=&  (\omega_m + Z_m b_m^\dag b_m) b_m - D - b_r \nonumber \\
    i \frac{d }{ d\tau} b_r(z,t) &=&  (\omega_r + Z_r b_r^\dag b_r) b_r - b_m,
\end{eqnarray}
where the dimensionless parameters are $Z_i = -U_0 \kappa_i /
\hbar L \Delta_{mr} $, $\omega_i = (\mu - \Lambda_i)/\Delta_{mr}$,
$D = \Delta_{lm} \sqrt{L n_l}/\Delta_{mr}$, and $L$ is the length
of the transistor.  The destruction operators are normalized to
$b_i' = b_i / \sqrt{L}$, and the primes have been dropped. Since
the equations for each "slice" in $z$ are of the same form as Eq.
(\ref{eqn:SQ_bMbL_equations}), the analysis of Sec.
\ref{sec:3well_structure} is  valid for each "slice" separately.

As with the case of a trapped atom transistor, we take $Z_m D^2 = 1$
and use the fact that the largest tunneling rate corresponds to $n_m
\approx D^2$.  We refer to this as the saturation density, since
$n_m$ is the normalized density of atoms and not the total number as
it was with a trapped atom device.   Next, we assume that the middle
waveguide can be described by the potential
\begin{equation}
V(\textbf{r}_\perp) = \frac {1} {2} m \omega^{2}_\perp r_\perp^2,
\end{equation}
where $\omega_\perp$ is the transverse frequency of the guide.
The overlap integral associated with this potential is
\begin{equation}
\kappa_m = \frac {1} {2 \pi} \frac {1} {a_\perp^2}, \label{eqn:overlap2d}
\end{equation}
where $a_\perp$ is the transverse oscillator length and $a_\perp =
\sqrt{\hbar / m \omega_\perp}$. Using Eq. (\ref{eqn:Zm}) and
(\ref{eqn:overlap2d}), we evaluate the coupling frequency between
the middle and right guides as
\begin{equation}
|\Delta_{mr}| \approx a_s \omega_\perp \frac {n_m}{L}.
\label{eqn:WG_tun_freq}
\end{equation}

In terms of the velocity of the atoms $v$ and the flux entering
the middle guide $\Phi_{m}$ the density can be expressed as $n_m/L
= \Phi_{m}/v$, and Eq. (\ref{eqn:WG_tun_freq}) takes the form
\begin{equation}
|\Delta_{mr}| \approx a_s \omega_\perp \frac {\Phi_{m}} {v}.
\end{equation}
Assuming that the guide has a transverse frequency of 10~kHz, the
velocity 5~cm/sec and the saturation flux is $10^5$~atoms/sec, we
can evaluate the coupling frequency between the right and middle
guide as
\begin{equation}
\Delta_{mr} \approx 2 \pi \times 10^2 rad / sec.
\end{equation}
The dimensional switching time is
\begin{equation}
 t \approx 2 \times 10^{-1} sec,
\end{equation}
and the length of the device is
\begin{equation}
L \approx 1 cm.
\end{equation}
This length can be decreased by slowing the velocity of the atoms,
increasing the saturation flux or increasing the transverse
frequency of the waveguide.

With the above numbers, a change in the input flux of the middle
guide from $0.4 \times 10^5$ atoms/sec to $0.8 \times
10^5$~atoms/sec results in a change of flux in the output of the
right guide from $10^6$ atoms/sec to about $10^7$~atoms/sec.

To summarize, we have presented a theoretical analysis of a Bose
Einstein condensate in a nonsymmetric three-well potential which
shows transistor-like behavior.  We demonstrated the control of
atomic population in the right well by the population in the middle
well with an absolute and differential gains considerably larger
than one. The second-quantization formalism was then used to
evaluate the quantum-mechanical uncertainty due to a finite number
of atoms and extend the mean-field results to the limit of a small
number of atoms.

The BEC transistor can turn out to be useful in precision
measurements. The number of atoms that tunnel from the source to the
drain is very sensitive to the number of atoms in the gate. This
fact can be used to detect and amplify small changes in the number
of atoms in the gate. A waveguide based transistor is capable of
operating continuously and can be used to measure time-dependent
phenomena.  Applications of this device may include measurement of
inertial changes and electromagnetic fields. It is possible to
envision potentially more interesting applications by combining
several such devices so that, e.g., the amplified output of the
first transistor serves as control for the second.

\section{Acknowledgements}
This work was supported by the Defense Advanced Research Projects
Agency's Defense Science Office through a PINS program (Grant No.
W911NF-04-1-0043) and the Air Force Office of Scientific Research
(Grant No. FA9550-04-1-0460).


\end{document}